\documentclass[11pt,oneside]{amsart}

\theoremstyle{plain}    

\numberwithin{equation}{section} 
\numberwithin{figure}{section} 

\newtheorem{dfn}{Definition}[section]

\newtheorem{fact}{Proposition}[section]

\def\commut#1#2{\left[{#1},\,{#2}\right]}

\def\const{\mathop{\rm const}\nolimits}

\def\bar{\overline}

\def\d{\partial}

\newcommand{\dl}[1]{\frac{{\d}}{\d #1}}

\def\half{\frac{1}{2}}

\def\manL{{\mathcal L}}
\def\manM{{\mathcal M}}

\def\manX{{\mathcal X}}

\font\frbig=eufm10 scaled\magstephalf
\font\frscr=eufm10
\font\frscrscr=eufm8
\newfam\frfam
\textfont\frfam=\frbig
\scriptfont\frfam=\frscr
\scriptscriptfont\frfam=\frscrscr

\def\cP{{\mathcal P}}

\newcommand{\p}[1]{\mathsf{p}(#1)}

\def\PRD{Phys.\ Rev.\ D}
\def\NPB{Nucl.\ Phys.\ B}
\def\PLB{Phys.\ Lett.\ B}
\def\MPLA{Mod.\ Phys.\ Lett.\ A}
\def\CMP{Commun.\ Math.\ Phys. \,}
\def\IJMPA{Int.\ J.\ Mod.\ Phys.\ A}
\def\JMP{J.\ Math.\ Phys.\, \,}

\newcommand{\func}[1]{{{\mathcal F}_{#1}}}

\newcommand\Vect[1]{{\rm Vect}_#1}

\def\Z{{\rm Z}}
\def\Q{{\rm Q}}

\begin{document}
\parindent=18pt
\addtolength{\baselineskip}{3pt}
\addtolength{\parskip}{4pt}

\title[Zero locus reduction]{Zero Locus Reduction of the BRST
  Differential}

{\hfill{\lowercase{\tt hep-th/9906209}}\\[12pt]

\author[MA\,Grigoriev]{M.~A.~Grigoriev}
\address{Tamm Theory Division, Lebedev Physics Institute, Russian
  Academy of Sciences}

\begin{abstract}
  I point out an unexpected relation between the BV
  (Batalin--Vilkovisky) and the BFV (Batalin--Fradkin--Vilkovisky)
  formulations of the same pure gauge (topological) theory.  The {\it
    nonminimal sector\/} in the BV formulation of the topological
  theory allows one to construct the Poisson bracket and the BRST
  charge on some Lagrangian submanifold of the BV configuration space;
  this Lagrangian submanifold can be identified with the phase space
  of the BFV formulation of the same theory in the {\it minimal\/}
  sector of ghost variables.  The BFV Poisson bracket is induced by a
  natural even Poisson bracket on the stationary surface of the master
  action, while the BRST charge originates from the BV gauge-fixed
  BRST transformation defined on a gauge-fixing surface.
  The inverse construction allows one to arrive at the BV formulation
  of the topological theory starting with the BFV formulation.  This
  correspondence gives an intriguing geometrical interpretation of the
  nonminimal variables and clarifies the relation between the
  Hamiltonian and Lagrangian quantization of gauge theories.

  \noindent
  This is an extended version of the talk given at the QFTHEP-99
  workshop in Moscow, May 27 -- June 2,~1999
\end{abstract}

\maketitle

\section{Introduction}
The aim of this talk is to outline an unexpected relation between the
Batalin--Vilkovisky \cite{[BV]} quantization and its Hamiltonian
counterpart known as the the Batalin--Fradkin--Vilkovisky quantization
\cite{[BFV]}.  I investigate the pure gauge sector of a gauge theory.
The relevant models are the topological theories, i.e.  the theories
all of whose degrees of freedom are gauge ones.  To anticipate the
result, the correspondence is based on the zero locus
reduction~\cite{[AKSZ],[GST]}, i.e., the construction of an even (odd)
Poisson bracket on the zero locus of the odd nilpotent vector field on
the (anti)symplectic manifold.

A geometrical counterpart of the BV quantization configuration space
(see~\cite{[BT]}--\cite{[Bering]} for the geometrically covariant
formulation of the BV formalism) is the QP manifold (the manifold
equipped with an antibracket and a compatible odd nilpotent vector
field $\Q$; in the BV context, $\Q$ is usually given by
$\Q=(S,\cdot\,)$, where $S$ is the BV master action satisfying the
master equation $(S,S)=0$).  As shown in~\cite{[AKSZ],[GST]}, the zero
locus of the odd vector field $\Q$ on a QP manifold is an even
symplectic manifold, provided the appropriate nondegeneracy condition
is imposed on $\Q$.  In the BV context, the zero locus $\Z_\Q$ of
$\Q=(S,\cdot\,)$ is the stationary surface of the master action~$S$.

On the other hand, the construction proposed in \cite{[SZ],[DN]},
\begin{equation}
  \label{eq:Poisson}
  \{f,g\}=(f,(S,g))\,,\qquad  (S,S)=0\,,
\end{equation}
expresses the Grassmann-even operations through the antibracket and the
solution $S$ to the master equation.  It was shown there that
\eqref{eq:Poisson} is a Poisson bracket provided one defines it on the
subalgebra of functions that are commutative w.r.t. the antibracket.
Moreover, the binary operation \eqref{eq:Poisson} induces a Lie
algebra structure on the space of gauge symmetries in the BV
formalism~\cite{[SZ]} (see also~\cite{[GST]}).  It was also observed
in~\cite{[GST]} that~\eqref{eq:Poisson} is a well defined operation in
the algebra of smooth functions on the zero locus~$\Z_\Q$ considered
as the quotient of all functions modulo those vanishing on~$\Z_\Q$.

An important observation of \cite{[BM-QA],[BM-dual]} is
that~\eqref{eq:Poisson} has an even counterpart.  Indeed, one can
consider an even Poisson bracket and an odd nilpotent vector field
that preserves the Poisson bracket; then the
structure~\eqref{eq:Poisson} with the antibracket replaced by the even
Poisson bracket determines an antibracket similarly to
how~\eqref{eq:Poisson} determines a Poisson bracket.  The zero locus
of the odd nilpotent Hamiltonian vector field on an even symplectic
supermanifold is therefore endowed with an
antibracket~\cite{[GST2]}.\footnote{Another interesting feature of the
  even analogue of~\eqref{eq:Poisson} is that it expresses the odd
  bracket through the even one and therefore, quantizing the even
  bracket one can arrive at the quantum counterpart of the
  antibracket~\cite{[BM-QA]}.  The quantization amounts to replacing
  the Poisson brackets in~\eqref{eq:Poisson} with the commutator.  We
  do not discuss this very interesting subject here, however.}

Since the zero locus itself is an (odd) Poisson (super)manifold, one
can try to make it into a QP manifold.  {\it However, if we start with
a general QP manifold}, there are no additional structures that
could induce an odd nilpotent vector field {\it on the zero locus\/}
(since $\Q$ vanishes on $\Z_\Q$, it induces only the trivial
$\Q$-structure on $\Z_\Q$). A possible way to make $\Z_\Q$ into a QP
manifold is to solve the equation
\[
(S,S)=0\quad \p{S}=0, \quad \text{or} \quad
(~~\{\Omega,\Omega\}=0\quad 
\p{\Omega}=1~~)
\]
in either the odd or the even case; this is the generating equation
for the odd nilpotent vector fields on $\Z_\Q$.  In this way, one
arrives at various relations between different QP~manifolds.  This
approach is developed in~\cite{[GST2]}, where it is also interpreted
in terms of Lie algebra (co)homology.

My claim is that there exists an alternative way to equip the zero
locus of $\Q$ with an odd nilpotent vector field.  This approach is
more restrictive but more relevant as regards applications in the BRST
quantization.  Namely, it can be possible to reduce to the zero locus
not only the bracket structure but also the BRST differential.  This
reduction can be viewed as a procedure relating the BV configuration
space and the BFV phase space of the same pure gauge model; the~BV
antibracket and the BV-quantization BRST differential
$\Q=(S,\cdot\,)$ induce the BFV phase space structures (the Poisson
bracket and the corresponding BRST differential
$\Q_{BFV}=\{\Omega,\cdot\,\}$) on the zero locus of 
$\Q=(S,\cdot\,)$.  In what follows, I will refer to this relation as
the \textit{zero locus reduction of the BRST differential}.

\section{QP-manifolds}
The notion of the QP manifold was introduced
in~\cite{[ASS-BVgeometry],[AKSZ]} as the geometrical structure
underlying the BV quantization.  In fact, a similar structure (with
the antibracket replaced by the ordinary Poisson bracket) underlies
the BFV quantization.  I now give a unified definition.
\begin{dfn}\label{def:QP}
  A QP~manifold is a supermanifold $\manM$ equipped with a
  nondegenerate antibracket (Poisson bracket) $({}\cdot{},{}\cdot{})$
  and an odd nilpotent vector field $\Q$ satisfying
  \begin{equation}\label{Leibnitz}
    \Q(F,\,G)
    -(\Q F,\,G)-(-1)^{\p{F}+\p{({}\cdot{},{}\cdot{})}}(F,\,\Q G)=0\,,
    \quad F,G\in\func\manM\,,
  \end{equation}
with $\func\manM$ being the algebra of smooth functions on $\manM$
and $\p{\cdot\,}$ being the Grassmann parity.
QP~manifolds with a Poisson bracket are called even, and those with
an antibracket, odd.
\end{dfn}

Let $\manM$ be a QP manifold. Let also $\Z_\Q$ be the zero locus of
$\Q$ (the submanifold where $\Q$ vanishes); this is assumed to be a
smooth manifold and $\Q$ is required to be regular. This means that
the components of $\Q$ generates an ideal of functions vanishing on
$\Z_\Q$.  One then has
\begin{fact}\cite{[GST]}\label{lemma:binary}
  The binary operation
  $\{\,,\,\}~{\rm :}~\func{\Z_\Q} \times
  \func{\Z_\Q}  \to \func{\Z_\Q}$
  \begin{equation}\label{eq:ZQPB}
    \{f,g\} = (F,QG)|_{\Z_\Q}\,,\qquad
    f,g \in\func{\Z_\Q}\,,\quad
    F,G \in \func\manM\,,\quad F|_{\Z_\Q}=f\,,\quad G|_{\Z_\Q}=g\,.
  \end{equation}
  is well defined (i.e., is independent of the choice of the
  representatives $F,G$ of functions $f,g$ on $\Z_\Q$). Moreover,
  $\{\,,\,\}$ is a Poisson bracket (antibracket) on $\Z_\Q$.
\end{fact}

Among the QP manifolds, I single out those on which the $\Q$ vector
field is maximally nondegenerate in the following sense: the matrix
$(\dl{\Gamma^A}Q^B)|_{p\in Z_Q}$ of the linear operators
$\Q_p:T_p\manM \to T_p\manM$ labelled by $p\in\Z_\Q$ should be of the
maximal rank at each point $p\in\Z_\Q$.  Since $Q_p$ is nilpotent,
this condition is formalized as follows
\begin{dfn}\label{def:proper}
  A QP~manifold $\manM$ is called \emph{proper} if the homology of the
  linear operator $\Q_p:T_p\manM\to T_p\manM$ is trivial at each point
  $p\in\Z_\Q$.
\end{dfn}
In coordinate-free terms, one can define this linear operator $\Q_p$
as follows.  For $p\in\Z_\Q$, one considers the tangent space
$T_p\manM$ as the quotient of $\Vect\manM$ modulo vector fields from
$\Vect\manM$ that vanish at~$p$.  It is easy to see that the operation
\begin{equation}
  \Q_p (X|_p)=(\commut{\Q}{X})|_p\,,
\end{equation}
is well-defined provided $\Q$ vanishes at $p$.  This gives an
important property of the $\Q$ field on a proper QP manifold:
\begin{fact}
  Let $\manM$ be a proper QP manifold. Then each function $f$
  satisfying $\Q f=0$ can be represented as $f=\Q h +\const$ in an
  appropriately small neighborhood of any point $p\in \Z_\Q$.
\end{fact}
In different words, the cohomology of $\Q$ is trivial when evaluated
in an appropriately small neighborhood of any point $p \in \Z_\Q$.  If
in particular $\Q=(S,\,\cdot\,)$, it follows that
\begin{equation}\label{eq:primitive}
  S=({\bar \Omega},S)
\end{equation}
for some (locally defined) function ${\bar\Omega}$ (since $S$ is
defined up to a constant, the constant term is omitted here).

Another important property of proper QP manifolds is the following
\begin{fact}\cite{[AKSZ],[GST]}
  Let $\manM$ be a proper QP~manifold. Then $\Z_Q$ is
  (anti)symplectic. The corresponding Poisson bracket (antibracket) is
  given by~\eqref{eq:ZQPB}.
\end{fact}
\noindent
{\sc Remark.}  All the properties of QP manifolds stated above do hold
if the manifold is finite-dimensional.  In the infinite dimensional
case, however, one should apply these statements with some care.
In particular, the standard master action $S$ of e.g., the Yang-Mils
theory is a proper solution of the master equation (which is precisely
the condition for $\Q=(S,\cdot\,)$ to be proper) but the homology
of $\Q$ is not trivial when evaluated on the local\footnote{here,
\textit{local} refers to space-time locality.} functionals.

\section{Reduction of the BRST differential}
\label{sec:reduction}
Let us be given a topological theory described by the master action
$S$ satisfying the classical master equation:
\begin{equation}
  \label{eq:master}
  (S,S)=0\,.
\end{equation}

Gauge fixing in the BV framework is described by the anticanonical
transformation $\psi$:
\begin{equation}
  \label{eq:S-gf}
  S_{\Psi}=\psi S=e^{(\Psi ,\cdot\, )}S,
\end{equation}
where $\Psi$ is a gauge fermion.  The BRST transformation acting
on the field-antifield space is
\begin{equation}
  \Q =(S,\cdot\, ).
\end{equation}
I also need the BRST transformation $\Q _{gf}$ acting on the
gauge-fixing surface (which is identified with $\manL_{0}$ via
$\manL_{\psi}=\psi ^{-1}\manL_{0}$).  In the covariant BV approach,
however, there is no natural way to determine the gauge-fixed BRST
transformation.  Indeed, being defined on the whole field-antifield
space $\manM$, the BRST transformation does not restrict naturally to
the Lagrangian submanifold $\manL_\Psi \subset \manM$.  Thus, in order
to determine the gauge-fixed BRST transformation, one needs an
additional structure.

Let the field-antifield space $\manM$ be the odd cotangent bundle over
some ``field space'' $\manL_0$ (this is the case in various
applications of the BV quantization).  Let also $\phi^A$ and
$\phi^*_A$ be the coordinates on $\manL_0$ and on the fibers (fields
and antifields).  Then it is possible to determine the gauge-fixed
BRST transformation via
\begin{equation}
  \label{eq:BRST-gf}
  \Q_{gf}f=(S_{\Psi},\pi^{*}f))|_{\manL_0}\,,\qquad h\in\func{\manL_0}\,,
\end{equation}
where $\pi^*$ is the pullback associated with the canonical projection
$\pi \, {\rm :} \, \manM \to \manL_0$ (again, gauge fixing submanifold
$\manL_\Psi$ is identified with $\manL_0$
via $\manL_{\psi}=\psi^{-1}\manL_{0}$).

Unfortunately, this prescription violates the nilpotency of the BRST
transformation in general.  An important case where $\Q_{gf}$ is still
nilpotent is where $S$ is linear in the antifields.  
I will thus assume that $S$ has the form
\begin{equation}
  \label{eq:linear-master}
  S=S_0(\phi)+\phi^{*}_{A}S^{A}\,.
\end{equation}
Now the gauge-fixed BRST transformation can be represented as
\begin{equation}
  \label{eq:BRST-gf-linear}
  \Q_{gf}h=(\Q (\pi ^{*}h))|_{\manL_{0}}=
(-1)^{\p{\phi^A}} S^{A} \dl{\phi^{A}}\,,
\end{equation}
and does not, therefore, depend on the particular choice of the gauge
fermion.

Let $\Z_{\Q}$ be the zero locus of the odd vector filed
$\Q=(S,\cdot\, )$ (the stationary surface of the master action).
Assume that there exists an anticanonical transformation $\varphi$
such that
\begin{itemize}
\item $\varphi (\manL_0)=\Z_{\Q}$, i.e., $\varphi$ transforms
  the initial Lagrangian submanifold $\manL_0$
  into $\Z_{\Q}$;
  
\item $\varphi$ is such that
  \begin{equation}
    \label{eq:TCC}
    S=(\varphi^{*}S,\Omega)
  \end{equation}
  for some new (in general, locally defined)
  function $\Omega\,$; here $\varphi^*$ is
  the pullback associated to~$\varphi$.  This
  can be expressed by saying that transformation
  $\varphi$ preserves the property of $S$ to be $Q$-closed.
\end{itemize}

The symplectic structure on $\Z_S$ is carried over to $\manL_0$ by the
anticanonical transformation $\varphi$.  The corresponding Poisson
bracket on $\manL_0$ reads
\begin{equation}
  \label{eq:PB}
  \{f,g\}=(\pi ^{*}f,(\varphi ^{*}(S),\pi ^{*}g))|_{\manL_{0}}\,,
  \qquad 
  f,g\in \func{\manL_0}\,.
\end{equation}

Let me now return to master equation~\eqref{eq:master} and rewrite
it in terms of the transformed action $\varphi^* S$ as
\[
((\varphi^*S,{\Omega}),(\varphi^* S,{\Omega}))=0\quad
\Longrightarrow \quad
(\varphi^*S,({\Omega},(\varphi^* S,{\Omega})))=0\,.
\]
The expression $(\Omega,(\varphi^* S,\Omega))$ formally coincides with
the Poisson bracket~\eqref{eq:PB} of $\Omega$ with itself.  To make
this correspondence manifest, observe that the equality
$(\varphi^*S,X)=0$ implies that (locally) $X=(\varphi^* S,Y)+\const$
and, thus, $X|_{\manL_0}={\rm const}$ provided $\manL_0$ is the zero
locus of~$(\varphi^* S,\cdot\,)$.  Now
$({\Omega},(\varphi^*S,{\Omega}))|_{\manL_0}=\const$ implies that
$({\Omega},(\varphi^*S,{\Omega}))|_{\manL_0}=0$, since the function
$\Omega$ is Grassmann-odd.  Thus,
\begin{equation}
  (\varphi^* S,({\Omega},(\varphi^* S,{\Omega})))=0\,, \quad
  \Longrightarrow 
  \quad \{ \Omega|_{\manL_0} , \Omega|_{\manL_0} \}=0\,,
\end{equation}
where $\{\cdot\,,\cdot\,\}$ is the Poisson bracket \eqref{eq:PB} on
$\manL_0$.

The crucial observation is that $\Q _{gf}$ considered as the vector
field on $\manL_{0}$ is generated by the bracket~\eqref{eq:PB}.
Indeed, Eq.~\eqref{eq:TCC} implies that
\[
\Q _{gf}h=(S,\pi ^{*}h)|_{\manL_{0}}=
(( \Omega , \varphi^{*}S),\pi^{*}h)=
\{\Omega|_{\manL_{0}},h\}\,,\qquad h\in\func{\manL_0}\,,
\]
where \( \{\cdot\, ,\cdot\, \} \) is the even Poisson
bracket~\eqref{eq:PB}.

Thus, the BV formulation of the topological theory gives rise to all
the data of the corresponding BFV scheme, including the Poison bracket
and the generating function $\Omega$ satisfying $\{\Omega,\Omega\}=0$.
The corresponding BRST transformation $\{\Omega,\cdot\,\}$ coincides
with $\Q_{gf}$ in the BV framework.

I have to note, however, that this coincidence is rather formal: the
main difference between the BV and BFV pictures is that in the BV
framework, one works with the \textit{configuration space} (the space
of field histories), while in the BFV approach one considers the
\textit{phase space}.  Further, the BV antibracket is defined on the
functionals on {\it space-time}, and the BFV Poisson bracket is the
equal-time bracket defined on the functionals on the space.  In
Sec.~\ref{sec:real}, I will give some arguments that make this
coincidence exact in several special cases.

\section{Existence of $\varphi$}

The issue of the existence of a globally defined
transformation~$\varphi$ is very complicated in the general setting.
I will limit myself to a local analysis.  The argument is based on
Abelianization.  It is well known that given a proper solution $S$ to
the master equation, one can find new canonical coordinates in some
(appropriately chosen) neighborhood, where $S$ becomes
quadratic.  Specializing this to the case where $S$ has the form
$S=\phi^*_A S^A(\phi)$
(since the theory is topological one can assume that the initial
action $S_0$ from \eqref{eq:linear-master} vanishes)
in the initial coordinates, one arrives at the
new coordinates $z^\alpha,y^\alpha,z^*_\alpha,y^*_\alpha$ such that
\begin{equation}\label{eq:minimal-abelian-master}
  S=z^*_\alpha y^\alpha\,,\qquad
  (z^\alpha,z^*_\beta)=\delta^\alpha_\beta\,,\quad
  (y^\alpha,y^*_\beta)=\delta^\alpha_\beta\,,
\end{equation}
and the submanifold $\manL_0$ is determined by the equations
$z^*_\alpha=y^*_\beta=0$. It is easy to see that in the new
coordinates, the stationary surface $\Z_{\Q}$ is determined by the
equations $y^\alpha=z^*_\beta=0$.

One can see that the transformation $\varphi$ relating the
submanifolds $\manL_0$ and $\Z_\Q$ does not exist in general; the
existence of $\varphi$ implies that the superdimensions of $\manL_0$
and $\Z_\Q$ are equal.  This in turn implies that among the variables
$x^\alpha$ and $y_\alpha$, there are equally many even and odd ones.
This is also a sufficient condition for $\varphi$ to exist, at least
locally.

Indeed, let me assume that the condition is fulfilled.  Then the
variables $z,z^*,y,y^*$ split into the pairs $[x^i,{\bar c}_i]\,,
[x^*_i,{\bar c}_*^i]\,, [c^i,b_i]\,, [c^*_i,b_*^i]$ of different
parities.  Now the master action from
\eqref{eq:minimal-abelian-master} takes the form
\begin{equation}
  S=x^*_ic^i+{\bar c}^i_* b_i\,.
\end{equation}
And the transformation, $\varphi$ is obviously given by
\begin{equation}
  \label{eq:abelian-phi}
  \begin{array}{c}
    \varphi^{*} ( x^{i} )=x^{i}\,\quad
    \varphi^{*} ( x^{*}_{i} )=x^{*}_{i}\,\quad
    \varphi^{*}( \bar{c}_{*}^{i} ) = \bar{c}^i_{*}\,\quad
    \varphi^{*}( \bar{c}_i )=\bar{c}_i\,\\
    \varphi^{*}(y^i)=b_{*}^{i},\quad
    \varphi^{*}(b^i_{*} )=-y^{i},\quad
    \varphi^{*}(b_i )=y^{*}_{i},\quad
    \varphi^{*}( y^{*}_i )=-b_{i}\,.
  \end{array}
\end{equation}
This transformation evidently satisfies all the conditions of
Sec.~\ref{sec:reduction}.

An important case occurs if the master action is constructed via the
standard BV prescription for a pure gauge model.  Then the required
condition always holds provided the configuration space is enlarged by
the nonminimal sector (i.e., by the standard auxiliary variables
needed for gauge fixing).  As we will see in the example in the next
section, this is indeed the case with an irreducible pure gauge
theory.  Thus the nonminimal variables are precisely those which allow
one to construct the~$\varphi$ transformation.

\section{A simple example}\label{sec:example}
Here, I illustrate the constructions of Sec.~\ref{sec:reduction} in a
very simple example of a topological model which is constructed in the
BV framework as follows.  Let me start with a smooth manifold $\manX$
considered as the configuration space of a model with the vanishing
action.  That the initial action vanishes implies that all the degrees
of freedom are pure gauge ones.  Let $R_\alpha$ be the corresponding
gauge generators, which I assume for simplicity to be linearly
independent and form a closed algebra
\begin{equation}
  [R_{\alpha },R_{\beta }]=C^{\gamma }_{\alpha \beta }R_{\gamma }
\end{equation}
In local coordinates $x^i$ on $\manX$, one has $R_\alpha=R^i_\alpha
\dl{x^i}$.  That all degrees of freedom are gauge ones implies that
$R_\alpha$ considered as vector fields on $\manX$ constitute a basis
of the tangent space to $\manX$ at each point; equivalently,
$R^i_\alpha$ is a nondegenerate matrix.

According to the standard BV quantization prescription, I now
introduce the ghost variables \( c^{\alpha } \) for each gauge
generator \( R_{\alpha } \) and the antifields variables
\(x^{*}_{i},c^{*}_{\alpha }\).  These variables constitute the
so-called \textit{minimal sector} of the gauge theory. The
corresponding solution to the classical master equation is
\begin{equation}
  \label{eq:min}
  S_{min}=x^{*}_{i}R^{i}_{\alpha} c^{\alpha }+
  \frac{1}{2}c^{*}_{\gamma}C^{\gamma }_{\alpha \beta }c^{\alpha
    }c^{\beta }\,. 
\end{equation}
To fix the gauge, one should enlarge the minimal set of variables to
the \textit{nonminimal one}.  The nonminimal variables are the
antighosts \( \bar{c}_{\alpha } \), the auxiliary filed \( b_{\alpha }
\), and their conjugate antifields.  One should add also the
nonminimal terms to the master action. The master action becomes
\begin{equation}
  \label{eq:nonmin}
  S=S_{min}+\bar{c}_{*}^{\alpha }b_{\alpha }\,.
\end{equation}
which again is a proper solution to the master equation.  Now the
gauge fixing can be performed by choosing a gauge fermion $\Psi$.

Let me now concentrate on the zero locus $\Z_{\Q}$ of the master
action~\eqref{eq:nonmin}.  The submanifold $\Z_{\Q}$ is determined by
the equations
$x^{*}_{i}=c^{\alpha }=\bar{c}^{*}_{\beta}=b_{\gamma}=0$
and is a Lagrangian submanifold of the configuration space. The
Poisson bracket \eqref{eq:ZQPB} on $\Z_\Q$ reads
\begin{equation}
  \label{eq:PBL}
  \{ x^i , c^{*}_{\alpha} \}=R^i_\alpha \,, \qquad
  \{ c^*_\alpha , c^*_\beta \}=
  -C^\gamma_{\alpha \beta} c^*_\gamma\,, \qquad
\end{equation}
with all the other brackets vanishing.  Thus $\Z_\Q$ is equipped with
a nondegenerate Poisson bracket.  At the same time, the submanifold
$\manL_0 \in \manM$ is equipped with the odd nilpotent vector field
$\Q_{gf}$ (a gauge fixed BRST transformation
\eqref{eq:BRST-gf-linear}); in the local coordinates $x,\pi,c,{\bar
  c}$, it reads
\begin{equation}
  \label{eq:Qstructure}
  \Q_{gf}=c^\alpha R^i_\alpha \dl{x^i}-\half C^\gamma_{\alpha \beta}
  c^\alpha c^\beta \dl{c^\gamma}-\pi_\alpha \dl{{\bar c}_\alpha}\,.
\end{equation}

Let me now turn to the construction of the anticanonical
transformation $\varphi$ relating the Lagrangian submanifolds $\Z_\Q$
and $\manL_0$.  For the Abelian theory (i.e., with the structure
constants $C^k_{ij}$ vanishing), the required transformation is simply
an obvious modification of transformation~\eqref{eq:abelian-phi}.  In
the nonabelian case, however, one should add additional terms.
Consider the following generalization of \eqref{eq:abelian-phi}
\begin{equation}
  \label{eq:nonabelian-phi}
  \begin{array}{c}
    \varphi^{*} ( x^{i} ) = x^{i}\,\quad
    \varphi^{*} ( x^{*}_{i} ) = x^{*}_{i}\,,\quad
    \varphi^{*}(c^{\alpha}) = b_{*}^{\alpha }\,,\quad
    \varphi^{*}(b^{\alpha}_{*} )=-c^{\alpha }\,, \\
    \varphi^{*}(b_{\alpha} )=c^{*}_{\alpha }-C^\gamma_{\alpha \beta}
    {\bar c}_\gamma b_*^\beta \,,\quad
    \varphi^{*}(c^{*}_{\alpha} )=-b_{\alpha }-C^\gamma_{\alpha \beta}
    {\bar c}_\gamma c^\beta\,, \\
    \varphi^{*}( \bar{c}_{\alpha} )=\bar{c}_{\alpha }\,, \quad
    \varphi^{*} (\bar{c}_{*}^{\alpha } )=\bar{c}^{\alpha }_{*}
    +C^\alpha_{\gamma \beta} b_*^\gamma c^\beta \,.
  \end{array}
\end{equation}
We have
\begin{fact}
  Transformation \eqref{eq:nonabelian-phi} is anticanonical (preserves
  the antibracket), is globally defined, and satisfies the condition
     \begin{equation}\label{eq:Tcond}
    S=(\Omega , \varphi^* S)\,.
     \end{equation}
\end{fact}

Indeed, transformation \eqref{eq:nonabelian-phi} is globally defined
by construction (since it preserves the base~$\manX$).  The
transformed master action is
\begin{equation}
  \varphi^{*}S=x^{*}_{i}R^{i}_{\alpha }b_{*}^{\alpha}-
  \half b_\gamma C^\gamma_{\alpha \beta} b_*^\alpha b_*^\beta
  +{\bar c}_*^\alpha c^*_\alpha
  -{\bar c}_*^\alpha C^\gamma_{\alpha \beta} {\bar c}_\gamma b_*^\beta
  -c^*_\gamma C^\gamma_{\alpha \beta} c^\beta  b_*^\alpha\,.
\end{equation}
It is easy to see by direct computations that
\begin{equation}
  \label{eq:nonabelian-omega}
  \Omega= c^\alpha b_\alpha - \half {\bar c}_\gamma C^\gamma_{\alpha
    \beta} c^\alpha c^\beta \,.
\end{equation}
solves the condition \eqref{eq:Tcond}.

It now follows from the general treatment of section
\ref{sec:reduction} that $\manL_0$ is equipped with Poisson
bracket~\eqref{eq:PB}.  In the local coordinates on $\manL_0$, we have
\begin{equation}
  \label{eq:nonabelian-PB}
  \begin{split}
    \{x^{i},b_{\alpha }\}=-R^i_\alpha \,,\quad
    \{b_\alpha,b_\beta\}=C^\gamma_{\alpha \beta} b_\gamma \,, \\
    \{\bar{c}_{\alpha },c^{\beta }\}=-\delta^{\beta }_{\alpha}\,,\quad 
    \{\bar{c}_{\alpha }, b_\beta \} =
    C^\gamma_{\alpha \beta} \bar{c}_{\gamma} \,, \quad 
\{ c^{\alpha} , b_\beta \}=C^\alpha_{\beta \gamma} c^\gamma\,.
  \end{split}
\end{equation}
The Jacobi identity for bracket~\eqref{eq:nonabelian-PB} holds
provided $\varphi^* S$ satisfies the classical master equation.

It also follows from the results of Sec.~\ref{sec:reduction} that
$\Q_{gf}$ given by \eqref{eq:Qstructure} is compatible with Poisson
bracket~\eqref{eq:nonabelian-PB}.  Moreover,
\begin{equation}
Q_{gf}=\{\Omega,\cdot\,\}
\end{equation}

Thus, at least at the formal level, we obtain all the objects of the
BFV quantization: the Poisson bracket and the generating function (the
BRST charge) $\Omega$.  To make the coincidence more clear let me
choose new coordinates
\begin{equation}
  q^i=x^i\,, \quad
  p_i=-R^\alpha_i (b_\alpha+C^\gamma_{\alpha \beta}{\bar c}_\gamma
  c^\beta)\,,\quad 
  {\bar \cP}_\alpha=-{\bar c}_\alpha\,, 
\end{equation}
with all the others being unchanged.  These are the canonical
(Darboux) coordinates for the Poisson bracket~\eqref{eq:nonabelian-PB}
\begin{equation}
  \{x^{i},p_j\}=\delta^i_j\,,\quad
  \{c^{\alpha },{\bar \cP}_\beta\}=\delta_{\beta }^{\alpha }\,.
\end{equation}
In the new coordinate system, one gets
\begin{equation}
  \Omega=c^\alpha T_\alpha - 
  \half {\bar \cP}_\gamma C^\gamma_{\alpha \beta}c^\alpha
  c^\beta\,,\qquad 
  T_\alpha=-R^i_\alpha p_i\,.
\end{equation}
which is the standard form of the BRST charge for the pure gauge
theory with the first-class constraints given by $T_\alpha$.
Moreover, at the formal level,\footnote{An account is not yet taken of
  an explicit time dependence of the Hamiltonian scheme.  This issue
  will be explained in Sec.~\ref{sec:real}} this is the BRST charge of
the same theory. Indeed, the first class constraints $T_\alpha$
satisfy the same Lie algebra as the gauge transformation $R_\alpha$ in
the BV formulation; the phase space of this system is the cotangent
bundle $T^*\manX$ enlarged by the ghosts $c^\alpha$ and the
corresponding momenta ${\bar \cP}_\alpha$.

A remarkable feature of the proposed reduction procedure is that it
reduces the BV scheme of the topological model in the {\it
  nonminimal\/} sector to the corresponding BFV scheme in the {\it
  minimal\/} sector.  Moreover, the zero locus reduction of the BRST
differential identifies the nonminimal variables in the BV
quantization with the momenta associated to the minimal variables in
the BFV picture of the same theory.

\section{A Generalization}\label{sec:real}

Until this point, I discussed a finite dimensional analogue of the
topological theory in the BV and BFV framework.  I would now like to
give some arguments showing that the zero locus reduction works well
not only at the formal level.  For simplicity, I consider only the
1-dimensional field theory, or quantum mechanics.  To avoid
considerable technical complications, let me also assume that the
topological theory at hand can be represented in the form where the
master action and the gauge generators come from the corresponding
objects in the target space.  This means that
\begin{equation}
  \label{eq:aksz-master}
  S=\int dt~s(\phi(t),\phi^*(t))\,,
  \qquad
  R^i_\alpha(t,t^\prime)=\delta(t-t^\prime)r^i_\alpha(\phi(t))\,, 
\end{equation}
where $s$ and $r_\alpha$ are functions and vector fields on the target
space (which is also antisymplectic).  This is the case for a wide
class of pure gauge theories including e.g., SQM, Chern--Simons model, and
topological sigma models (see~\cite{[AKSZ]} for details).

Then the BV configurations space is the space of smooth maps from the
``time line'' into the target space.  The configuration space is
equipped with the antisymplectic structure given by the lift of the
antisymplectic structure in the target space to the space of maps into
the target space (see \cite{[AKSZ]}).

Since the zero locus reduction can be performed in the target space,
one arrives at the symplectic submanifold $\manL_0$ equipped with the
odd vector field $Q_{gf}=\{\Omega,\cdot\,\}$.  The space of maps from
the time line into the symplectic manifold can be identified with the
space of Hamiltonian trajectories.  Under this identification, the
Poisson bracket in the target space becomes the equal-time Poisson
bracket.  One thus arrives at the BFV formulation of the same
topological model.

\section{The BV quantization from the BFV approach.}
Similar considerations can be applied starting with the BFV
formulation of a topological model.  In that case, the zero locus of
the BFV BRST differential $\Q=\{\Omega,\cdot\,\}$ is equipped with a
nondegenerate antibracket. A similar reduction procedure (where one
should find the transformation relating the zero locus of $\Q$ and the
submanifold of coordinates) provides this submanifold with an odd
Hamiltonian vector field $Q$, which can be interpreted as the BRST
transformation generated by some (Grassmann-even) function $S$ viewed
as the master action in the corresponding BV formulation.  Again, the
BFV formulation in the nonminimal sector reduces to the BV formulation
in the minimal sector.

\section{Conclusions}
We have explicitly constructed the BFV phase space of a pure gauge
(topological) model starting with the BV configuration space of the BV
formulation of the same model.  The construction is based on two
important ingredients.

The first one is the even Poisson structure on the zero locus of the
BRST differential (stationary surface of the master action) in the BV
formalism~\cite{[AKSZ],[GST]}.

The second is the construction of the appropriate anticanonical
transformation $\varphi$ relating the gauge fixing surface (which in
our simplified approach is identified with the initial submanifold
$\manL_0$ via the gauge fixing transformation) and the stationary
surface $\Z_\Q$ of the master action.  The transformation $\varphi$
allows us to identify $\Z_\Q$ with $\manL_0$ and, thus, to carry over
the gauge fixed BRST transformation $\Q_{gf}$ defined on $\manL_0$ to
a Hamiltonian vector field on $\Z_\Q$ (equivalently, to carry over the
Poisson bracket from $\Z_\Q$ to $\manL_0$; this is the viewpoint
adopted in the paper).  In this way, one arrives at the Lagrangian
submanifold equipped with two compatible structures, the even Poisson
bracket and the odd nilpotent vector filed.  A further analysis shows
that one can formally identify this Lagrangian submanifold with the
phase space of the corresponding BFV phase space.

An important feature of the proposed construction is that the BV
variables of the nonminimal sector play the role of momenta that are
conjugate to the variables of the BFV minimal sector.  This gives an
interesting geometrical interpretation of the nonminimal
variables.

It should be noted that the relation discussed here is essentially
based on the fact that the initial BV structures correspond to the
topological theory.  However, in the general gauge theory
setting, one can formally separate the physical sector and the pure
gauge sector and then perform the reduction in the pure gauge sector.
This suggests an interpretation of the results presented here as those
describing the structure of the pure gauge sector of the general gauge
theory.

This also suggests that the Poisson bracket in the BFV formulation of
the general gauge theory splits into two parts: the standard Poisson
bracket in the physical sector and the Poisson bracket in the pure
gauge sector.  The latter is related to the BV antibracket from the
corresponding BV formulation via the present construction.  At least
for an irreducible theory with a closed gauge algebra, this relation
originates in the fact that both the BV antibracket in the pure gauge
sector and the BFV Poisson bracket in the pure gauge sector come from
a single Gerstenhaber-like bracket structure in the gauge algebra
complex (see~\cite{[GST2]} for the details).  This is somewhat
reminiscent of the results of~\cite{[Isomorphism]}, where the general
relationship was established between certain Lie algebras w.r.t. the
BV antibracket and their Hamiltonian counterparts w.r.t. the BFV
Poisson bracket.

\subsubsection*{Acknowledgments}
I am grateful to A.~M.~Semikhatov for his attention to this work and
numerous discussions and suggestions.  I also wish to thank
I.~A.~Batalin and O.~M.~Khudaverdian, and especially, I.~Yu.~Tipunin,
I.~V.~Tyutin and B.~L.~Voronov for illuminating discussions.  This
work was supported in part by the RFBR Grant 99-01-00980, INTAS-YSF
98-156 and the Russian Federation President Grant~99-15-96037.
I also acknowledge a partial support from the Landau Scholarship
Foundation, Forschungszentrum J\"ulich.

\end{document}